\def\year{2020}\relax
\documentclass[letterpaper]{article} 
\usepackage{aaai20}  
\usepackage{times}  
\usepackage{helvet} 
\usepackage{courier}  
\usepackage[hyphens]{url}  
\usepackage{graphicx} 
\usepackage{multirow}
\usepackage{graphicx}  
\usepackage{epstopdf}
\newcommand{\ie}{{\em i.e.,}\xspace}

\newcommand{\BfPara}[1]{{\noindent\bf#1.}\xspace}
\usepackage{tikz}
\usepackage{booktabs}
\newcommand*\cib[1]{\tikz[baseline=(char.base)]{
                            \node[shape=circle,color=black, fill=black!70!white,text=white,draw,inner sep=0.3pt] (char) {#1};}}
\newcommand{\citet}[1]{\citeauthor{#1} \shortcite{#1}}
\usepackage{comment}
\usepackage{blindtext}
\usepackage{booktabs} 
\usepackage{amsfonts}
\usepackage{amsmath} 
\usepackage{booktabs}
\usepackage{multirow}
\usepackage{graphics}
\usepackage{subfigure}
\usepackage{xcolor}
\usepackage{tikz}
\usepackage{enumitem}
\usepackage{booktabs} 
\usepackage{xspace}
\newcommand{\tma}{{\sf TM-1}\xspace}
\newcommand{\tmb}{{\sf TM-2}\xspace}
\newcommand{\tmc}{{\sf TM-3}\xspace}
\usepackage[utf8]{inputenc}
\usepackage{makecell}
\usepackage{amssymb,amsfonts,balance}
\usepackage{amsmath} 
\usepackage{graphicx}
\usepackage{subfigure}
\usepackage{wrapfig,lipsum,booktabs}
\usepackage{multirow}
\usepackage{color}
\usepackage{xcolor}
\usepackage{listings}
\usepackage{url}
\usepackage{mathenv}

\usepackage{multicol}
\usepackage{xspace}
\usepackage{setspace}
\usepackage{enumitem}
\usepackage{mathtools}

\begin{document}
\pdfinfo{
/Title (But You Cannot Hide: Using Elevation Profiles to Breach Location Privacy through  Trajectory Prediction)
/Author (Ülkü Meteriz, Necip Fazıl Yıldıran, and Aziz Mohaisen)
/Keywords (location privacy, privacy breach, fitness applications, data mining, natural language processing, machine learning, deep learning, multi-class classification)
}

\title{You Can Run, But You Cannot Hide: \\ Using Elevation Profiles to Breach Location Privacy through  Trajectory Prediction}

\author{Ülkü Meteriz, Necip Fazıl Yıldıran, and Aziz Mohaisen\\\em University of Central Florida}

\maketitle

\begin{abstract}
\begin{quote}
The extensive use of smartphones and wearable devices has facilitated many useful applications. For example, with Global Positioning System (GPS)-equipped smart and wearable devices, many applications can gather, process, and share rich metadata, such as geolocation, trajectories, elevation, and time. For example, fitness  applications, such as Strava and Runkeeper, utilize information for activity tracking, and have recently witnessed a boom in popularity. Those trackers have their own web platforms, and allow users to share activities on such platforms, or even with other social network platforms. To preserve privacy of users while allowing sharing, those platforms allow users to disclose partial information, such as the elevation profile for an activity, which supposedly will not leak the location trajectory.

In this work we examine the extent to which publicly available elevation profiles can be used to predict the location trajectory of users. To tackle this problem, we devise three threat settings under which the city, borough, or even a route can be predicted. Those threat settings define the amount of information available to the adversary to launch the prediction attacks. Establishing that simple features of elevation profiles, e.g., spectral features, are insufficient, we devise both natural language processing (NLP)-inspired text-like representation and computer vision-inspired image-like representation of elevation profiles, and we convert the problem at hand into text and image classification problem. We use both traditional machine learning- and deep learning-based techniques, and achieve a prediction success rate ranging from 59.59\% to 95.83\%. The findings are alarming, and highlight that sharing information such as elevation profile may have significant privacy risks.  

\end{quote}
\end{abstract}

\section{Introduction}

From smartphones to wearable devices, various types of Internet of Things (IoT) devices are equipped with Global Positioning System (GPS), accelerometers and gyroscopes to allow applications to function or to present a better user experience by making use of {\em geodata}, such as location and elevation information. Specifically, fitness applications which run on smartphones and smartwatches use these systems to collect spatial, temporal, and activity-specific information to analyze, summarize and visualize users' activities. By analyzing each activity, many of those applications even deliver personalized motivations and challenges for users to meet their goals. Using social media support of these applications for sharing updates about users' activities, including training routes and elevation profiles for the routes taken for the activity (e.g., walking, running, climbing, cycling), users can have positive behavioural changes through a more active lifestyle motivated by competitions with friends and acquaintances~\cite{higgins}.

Despite the broad set of advantages geodata has, geodata usage and uncontrolled sharing can pose a significant privacy risk which can be exploited in multiple attacks, including stalking \cite{Polakis:2015:WWP:2810103.2813605} and cybercasing \cite{Cybercasing}. For example, with the large amount of geo-tagged data, including text, images, and videos, cybercasing allows a significant attack vector to criminals and maliciously motivated individuals. Geo-tagged photos that are frequently posted on image sharing websites, such as Flickr, or second-hand sale websites, such as craigslist, may put owners of those images at risk. For example, geo-tagged images posted on sales websites may reveal the location of the advertised product, leading to trespassing or even theft. 
\begin{figure}[t]
    \centering
    \includegraphics[width=0.45\textwidth]{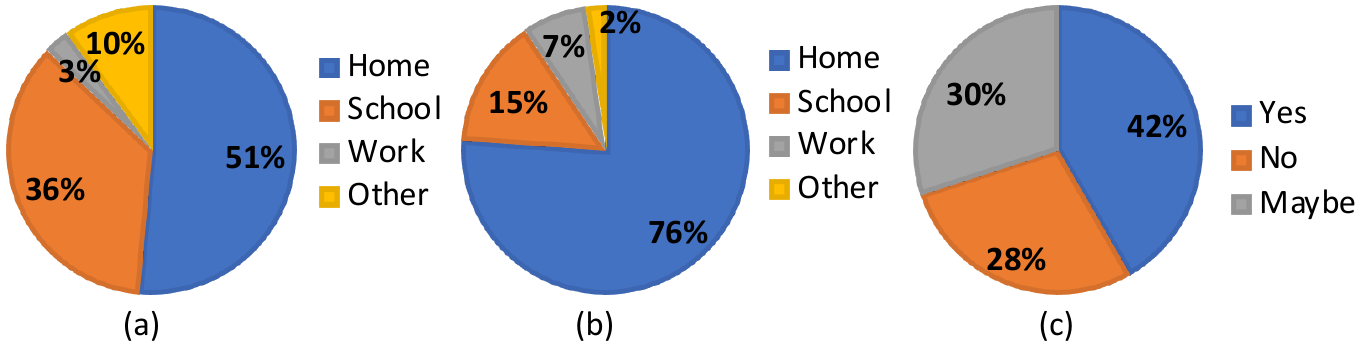}\vspace{-3mm}
    \caption{Survey results for understanding users behavior: (a) starting point statistics, (b) end point statistics, and (c) not sharing location information implies privacy. While 90\% of the 60 participants indicated their start of activity is either \textcolor{blue}{home}, \textcolor{orange!80!black}{school}, or \textcolor{gray!30!black}{work}, an overwhelming 98\% of the participant indicated those to be the end point of their activities.}
    \label{figure:survey}
\end{figure}
Geodata recorded by fitness applications are indeed important and valuable not only for the operation of those applications but also as an avenue for launching attacks on users by breaching their privacy, since sensitive information of users such as house or workplace location can be easily inferred from such data. To support this argument, we conducted a survey with 60 participants who regularly train outdoors, with the results of the survey summarized in Figure \ref{figure:survey}. The survey results reveal that 58.62\% of the participants start their training from their houses, 25.86\% start from their school and 3.45\% start from their workplace, and, almost 70\% of the participants finish their training at their houses. To protect such sensitive information, different fitness applications provide different mechanisms to keep parts of workout records private, as some of which could be seen from Table \ref{table:fitness_applications}. Although it is possible not to share exact route by using those mechanisms, people still may want to share elevation profile to show the roughness of the route they took as a measure of their workout. However, is the elevation profile of an activity enough to reveal some information about the activity? Interestingly, our survey results suggest that the surveyed subjects believe (or at least suspect, as with the ``maybe'' answers) that not sharing the exact location information (map trajectory) is sufficient to ensure privacy. In this paper, we argue that an approximate location at different levels of location granularity could still be revealed from elevation profile information, which are typically shared.

\BfPara{Contributions} In this paper, we contribute the following: 

\begin{itemize}[leftmargin=*]
    \item We translated the problem into text classification and image classification problems by encoding the elevation signals as strings and visualizing the elevation signals as images to employ the common approaches for solving image and text classification problems,
    \item We investigated the possible attack surface for the problem by introducing three different threat models, which we later used to evaluate the success of our approaches by simulating our methods considering each threat model,
    \item We proved that location information can be predicted from elevation profile using different machine/deep learning methods with accuracy in range 59.59\% - 95.83\% at different resolutions as our results showed.

\end{itemize}

\section{Background}
    \subsection{Fitness Applications}
    Fitness applications available for smartphones allow users to track their workout history and statistics. Some fitness applications have social network capabilities, as shown in Table~\ref{table:fitness_applications}, allowing users to share workout summaries, which is known to motivate users and their social network connections~\cite{higgins}. Moreover, among widely used fitness applications, Strava and Runkeeper have additional privacy settings for posts, allowing users to hide actual routes while sharing only the elevation profiles. Given its popularity, Strava is the application we consider in this work.
    
    \begin{table}[t]
        \caption{Fitness applications and their features, including exercise tracking (ET), ability to share to social media (SS), social networking capabilities in the service (SNS), privacy records (PR), and user blocking capability (BU).}
        \centering
        \begin{tabular}{cccccc}
             \toprule
            Service & ET & \makecell{SS} & SNS & PR & BU \\
             \midrule
             Strava         & \checkmark & \checkmark & \checkmark & \checkmark & \checkmark \\
             \hline
             Runtastic      & \checkmark & \checkmark & \checkmark &            & \checkmark \\
             \hline
             Runkeeper      & \checkmark & \checkmark & \checkmark & \checkmark & \\
             \hline
             Nike+ Running  & \checkmark & \checkmark & \checkmark & \checkmark & \\
             \hline
             MapMyRun       & \checkmark & \checkmark & \checkmark & \checkmark & \\
             \bottomrule
        \end{tabular}
        \label{table:fitness_applications}
    \end{table}
    
    \BfPara{Strava}
        Strava allows users to track and optionally share their outdoor workouts. Although the total number of users is unknown, Strava is expanding by 1 million users every 45 days and 8 million activities are recorded each week. In our study, we used the public Strava Development API to mine Strava Segments which are user-created training routes, and typically used for competition among users. The main advantages of segments are as follows. First, since segments are created for competition, the routes are frequently used by users, increasing the segment's likelihood of occurrence within a set of routes. Second, segments may include challenging terrain, possibly with significant change in elevation. We hypothesize that a rich profile of change in elevation within a route can be an identifying feature of such a route, which we explore and confirm in this work.

    \subsection{Privacy Breach Incidents in Strava} There has been a lot of privacy incidents concerning location data, particularly data obtained from Strava, which we review in the following to contextualize our work. 

\BfPara{Global Heatmap}
        Strava collects users' public data and publishes heatmap of the aggregates to highlight routes frequented by users. Although the aggregates in the heatmap do not explicitly contain any identity information, activities in desolate places revealed location of many U.S. military bases, which is sensitive information.
        
        \BfPara{Deanonymization Through Strava Segments}
        Strava heatmap provides anonymous public data. It is shown that a dedicated adversary can deanonymize heatmap to find out users who ran in a specified route~ \cite{AdvancedDeanonymizationThroughStrava}. For example, by selecting a route from the heatmap, a registered user can manually create a GPS eXchange (GPX) track file and a segment using it on Strava. Consequently, the top-10 users grouped by gender and age who previously ran that route are shown on the leaderboard, thus identifying them.
        
        \BfPara{Bicycle Theft} 
        Users on Strava can share their equipment including expensive bicycles and shoes, along with the routes frequented, making them a target to robbery. As a result, one victim  claimed losses of around 16,000 USD. 
        
        \BfPara{Attack on Privacy Zone}
        Privacy zones are used to obfuscate the exact starting and ending point of a route. In a recent study,  \citet{217618} were able to reveal the exact starting and ending point of a route that utilizes the privacy zone feature. They also claimed that around 95\% of the users are at risk of revealing their location information.
    
\section{Threat Models \& Approach Overview}
We outline the threat models under which this study is conducted. We describe three models under which the location privacy from publicly shared elevation profiles is breached. We then review our approach, including a pipeline that consists of data collection and preprocessing, feature extraction, and multi-class classification for location trajectory identification. We briefly discuss the phases of our pipeline, each of which is explained in details in implementation section.

\subsection{Threat Models} Our study utilizes three threat models: \tma, \tmb, and \tmc, which we outline below with their justification. The adversarial capabilities in \tma are greater than in \tmb and \tmc, making it more a restrictive (powerful) model. 

\BfPara{\cib{1} \tma} In \tma, we assume an adversary with some records of the workout history of a target user, and the goal of the adversary is to identify the location of the target user from the shared elevation profiles. \tma is justified by multiple plausible scenarios in practice. For example, such an adversary might have been a previous social network connection of the targeted user that was later blocked. In such a scenario, the adversary may have such previous workout records of the target from which the adversary may attempt to de-anonymize the target's activities. Another example might include group activities, in which two individuals (i.e., the adversary and target) may have shared the same route at some point. In either case, by knowing the target's history, the adversary's goal in this model is to identify recent whereabouts from publicly shared elevation profiles in workout summaries, thus breaching the target's privacy. 
 
\BfPara{\cib{2} \tmb} In \tmb, we assume an adversary with access to limited information such as the city in which the target lives. Such information is easily accessible from public profile summaries, \url{athlinks.com}, public records, etc. The adversary's goal in \tmb is to find out which region or part of a given city the target's activities are associated with. The \tmb use scenario may include a targeted user sharing private activities, in which the route is hidden while the elevation profile is shown. The adversary, knowing the city where the target lives, would want to identify the region (e.g., borough in the city) associated with the user's activity.  

\BfPara{\cib{3} \tmc} In \tmc, we assume an adversary trying to identify the target user's city using only publicly shared elevation profiles. We assume, however, the adversary has the ability to profile the elevation of cities, with information that is easily obtained from public sources (e.g., Google maps and Strava API). The use scenario of \tmc may be used as a stepping stone towards launching the attack scenario in \tmb upon narrowing down the search space to a city.

\subsection{Approach Overview}

\BfPara{Data Collection and Preprocessing} We collected three datasets with varying and rich characteristics, namely 1) raw Strava data collected from Strava's users, 2) mined Strava segments grouped at city-level, and 3) mined Strava segments grouped at borough-level. For the raw dataset, we collected bulk raw activities of Strava users and converted those activities to our intermediate format, the GPS Exchange Format (GPX). Then, we parsed the GPX files and manually labeled them according to the latitude and longitude information included within each file. For the second dataset, we mined segment routes from Strava using Strava development API specifying the location boundaries, \ie the class label of the mined data, and augmented each segment route with the corresponding elevation profiles obtained from Google Maps Elevation API. Finally, the borough-level dataset is constructed in a similar manner as in the city-level dataset. 
In preprocessing, the collected datasets are transformed into text-like and image-like representations. For text-like representation, we discretized the elevation signals and computed the minimum required \emph{word} size. We then created a mapping between each unique discrete value and a string. By mapping the string correspondents to the unique discrete values, we encoded the elevation profiles in text. Finally, we form a \emph{vocabulary} from the text sequences of each dataset using the $n$-grams. To obtain image-like representation, we converted the elevation profiles to a fixed-sized line graph where the x-axis stands for time and y-axis stands for the elevation values. The lines in the graphs are also colored to represent the elevation interval in which the elevation profiles range.

\BfPara{Feature Extraction} The classification algorithm operates on high quality and discriminative features, obtained from the representations of elevation profiles. For feature extraction, we utilize Natural Language Processing (NLP) and computer vision approaches. To employ NLP approaches, using previously obtained vocabulary, we represent each elevation profile as a feature vector based on the frequency of the vocabulary in the text-like representation (bag-of-words vector). To employ computer vision approaches, we utilize Convolutional Neural Network (CNN) over image-like representations. The optimal features of an image-like representation are efficiently extracted by the convolutional and pooling layers in the CNN architecture.

\BfPara{Multi-class Classification} We use various machine and deep learning models for classification including Support Vector Machine (SVM) and Random Forest Classification (RFC) as machine learning approaches, and Multi-Layer Perceptron (MLP) and Convolutional Neural Network (CNN) as deep learning approaches.

\section{Implementation Details}\label{sec:implementation}
The implementation details of data collection, preprocessing, feature extraction and multi-class classification are addressed in the following subsections. 

    \subsection{Data Collection and Preprocessing}
    In this study, we use three datasets: the raw dataset, the city-level dataset, and the borough-level dataset. The raw dataset is retrieved from Strava users who have duplicate samples from similar trajectories in their profile. The raw dataset offers a dense and thorough coverage for regions frequented by users; those regions are used as class labels. The city- and borough-level datasets are created from scratch, and do not have any duplicate samples with similar trajectories. Both provide a spatial and comprehensive coverage of cities and boroughs. After data collection, data is preprocessed according to the feature extraction methods. We provide details on each dataset and the preprocessing method below.
    \subsubsection{Data Collection}

\begin{figure}[t]
    \centering
    \includegraphics[width=0.5\textwidth]{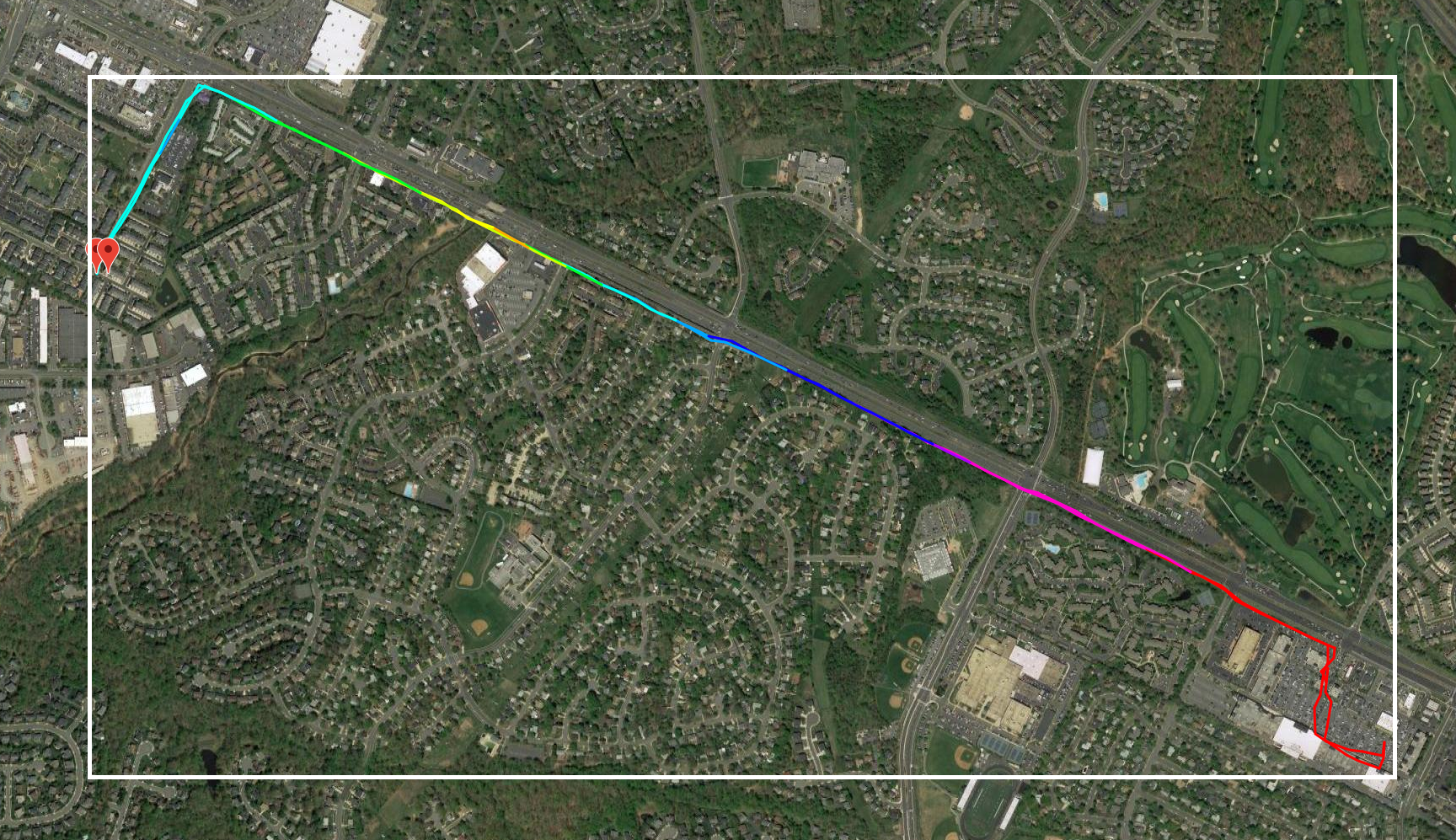}
    \caption{An illustration of the tight rectangle encapsulating an example route.}
    \label{figure:route}
    \end{figure}

    For the raw dataset, we collected data from Strava users whose activities are archived and can be downloaded through the Strava web interface. First, the raw data is converted to GPX format. For labelling, all samples are traversed, and the maximum and minimum coordinates of each sample are fetched. Each sample route is encapsulated with a tight rectangle whose top right (North East) and bottom left (South West) corners are computed from the maximum and the minimum coordinates of the route as illustrated in Figure \ref{figure:route}. To classify the samples, each rectangle encapsulating the route is compared with the previously created regions. If the Euclidean distance between the center of the rectangle and the center of the existing region does not exceed a predetermined threshold, the rectangle and its corresponding sample are labeled with a unique identity of the region. If there is no region includes the route, a new region is created. After the classification phase is done, we manually convert the labels from unique region IDs to region names by looking up where the coordinates of the regions are on world map.
    The final sample size distribution of raw dataset is shown in Table \ref{table:raw}.
    
\begin{figure}[t]
    \centering
        \includegraphics[width=0.5\textwidth]{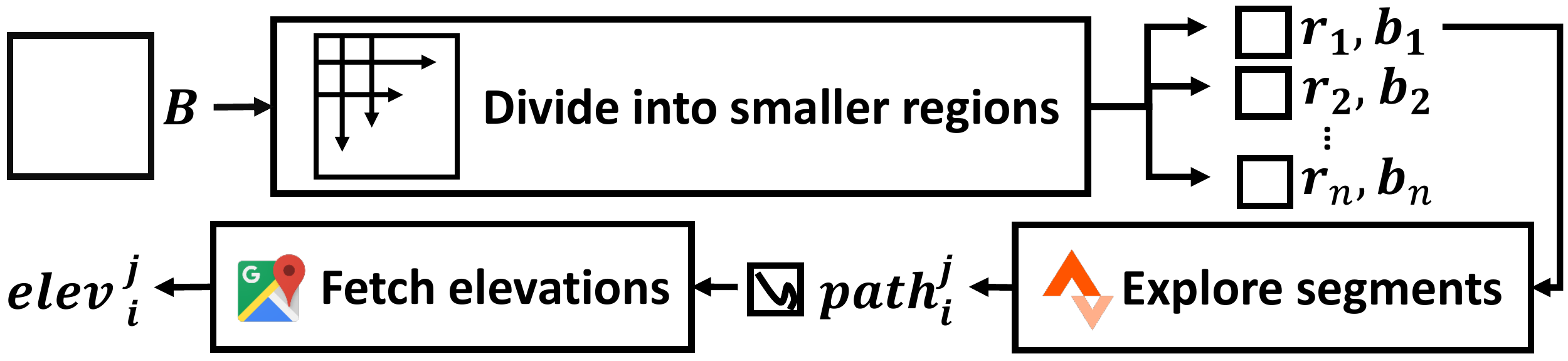}
        \caption{An illustration of data mining pipeline.}
        \label{figure:miningpipeline}
    \end{figure}

    For the city-level dataset, we mined publicly available Strava segments using the ``explore segments'' function of the Strava API. The overall procedure consists of 3 steps as illustrated in Figure \ref{figure:miningpipeline}. First, we defined the cities of interest, which are also the class labels. For each city $ C $, we defined rectangle boundaries, $ B $. Since the ``explore segment'' function returns the top-10 segments in a given boundary, we divided $ B $ into smaller regions, each denoted by $ r_i $, and computed their boundaries, each denoted by $ b_i $, to obtain more information about the city $ C $. Then, for each $ b_i $, we requested the segment information from the Strava API and received the geolocation polyline path, $path_i^j$ where $ j \in [1,10]$, of the top-10 segments included in $b_i$. Finally, we requested the elevation profile $ elev_i^j $ for each $path_i^j$ from the Google Maps Elevation API. The sample size distribution of city-level dataset can be found in Table \ref{table:city-level}.
    
    \begin{table}[t]
        \caption{Raw dataset sample size distribution.}
        \centering
        \begin{tabular}{lr}
             \toprule
             \textbf{Regions} & \textbf{Sample Size} \\
             \midrule
             Washington DC & 366 \\
             Florida & 232 \\
             New York City & 120 \\
             California & 18 \\
             \bottomrule
        \end{tabular}
        \label{table:raw}
    \end{table}
        
    \begin{table}[t]
        \caption{City-level dataset sample size distribution.}
        \centering
        \begin{tabular}{lr}
            \toprule
            \textbf{Regions} & \textbf{Sample Size} \\
            \midrule
            New York City & 2437 \\
            Washington DC & 2129 \\
            San Francisco & 743 \\
            Colorado Springs & 369 \\
            Minneapolis & 363 \\
            Los Angeles & 280 \\
            New Jersey & 266 \\
            Duluth & 156 \\
            Miami & 94 \\
            Tampa & 83 \\
            \bottomrule
        \end{tabular}
        \label{table:city-level}
    \end{table}
    
    For the borough-level dataset, we applied a similar mining procedure as at the city-level, using the borough boundaries instead of the city boundaries. Table \ref{table:borough-level} shows the sample size distribution of the borough-level dataset.

    \begin{table}[t]
        \caption{Borough-level dataset sample size distribution.}
        \centering
        \begin{tabular}{clr}
            \toprule
            \textbf{Cities} & \textbf{Regions} & \textbf{Sample Size} \\
            \midrule
            \multirow{4}{*}{ \makecell{\bf Los Angeles \\(LA)} }
            & Downtown & 280 \\
            & Santa Monica & 128 \\
             & Chinatown & 46 \\
            & Beverly Hills & 38 \\
            \midrule
            \multirow{3}{*}{\makecell{\bf Miami \\(MIA)}}
            & Downtown & 67 \\
            & Miami Beach & 44\\
            & Virginia Key & 18 \\
            \midrule
            \multirow{3}{*}{\makecell{ \bf New Jersey\\ (NJ) }}
            & Jersey City & 266 \\
            & West New York & 23 \\
            & Newark & 28 \\
            \midrule
            \multirow{6}{*}{\makecell{\bf New York City\\ (NYC)}}
            & Manhattan & 2437 \\
            & Queens & 353 \\
            & Brooklyn(South) & 239 \\
            & Brooklyn(North) & 205 \\
            & Bronx & 142 \\
            & Staten Island & 119 \\
            \midrule
            \multirow{4}{*}{\makecell{\bf San Francisco\\ (SF)} }
            & South West & 743 \\
            & South East & 144 \\
            & North West & 130 \\
            & North East & 86 \\
            \midrule
            \multirow{2}{*}{\makecell{\bf Washington DC\\ (WDC) }}
            & District of Columbia & 2129 \\
            & Baltimore & 218 \\
            \bottomrule
        \end{tabular}
        \label{table:borough-level}
    \end{table}

    \subsubsection{Preprocessing}
    We transform the samples to text- and image-like representations to facilitate feature extraction. 

\begin{figure}[t]
        \centering
        \includegraphics[width=0.5\textwidth]{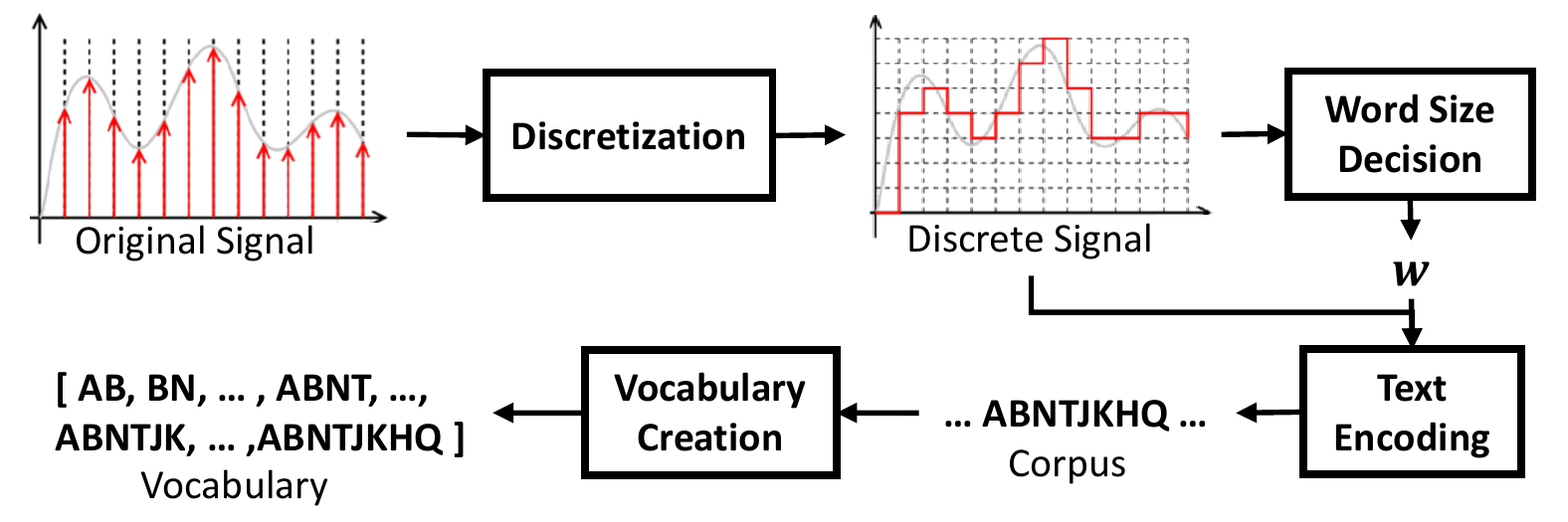}
        \caption{Illustration of the flow of text-like preprocessing. The signal is discretized by eliminating the small elevation fluctuations. The discretized signal is also used for deciding the word size of the encoding. The discrete signal is then encoded in text and a vocabulary is built.}
        \label{figure:text-preprocessing}
    \end{figure}

    \BfPara{Text-like Representation}     
    To represent data as text-like contents, we use four steps, as in Figure \ref{figure:text-preprocessing}:  discretization, word size decision,  text encoding, and vocabulary creation. 
    
    \cib{1} In the discretization step, the original elevation signal is discretized along the y-axis, which represents the elevation values to avoid possible overhead by small differences in the precision causing a longer string correspondences and, consequently, longer vocabulary and sparse feature vectors.  The discretization is done as follows. Let $e_i^j$ be the $i$-th elevation value in $j$-{th} sample. The discretization functions are defined as $f( e_i^j ) = \lfloor{} e_i^j \rfloor{}$ and $f( e_i^j ) = \frac{\lfloor{} e_i^j \times 10^3 \rfloor{}}{10^3}$, where the first function is used for processing the raw dataset and the second function is used for processing the city-level and borough-level datasets. Since the raw dataset is dense in terms of sampling rate, using the floor function is enough to represent the routes. However, as the city-level and borough-level datasets are already sparse, losing information is undesired, so we used the second function to represent the elevations that differ in up to 3 decimal digits precision. 
    
     \cib{2} For word size decision, we use $w = \log_l c$, where $w$ is the word size, $l$ is the length of the alphabet, and $c$ is the number of unique value occurrences in the given signals. 
     
     \cib{3} For text encoding, each unique value in all the discrete signals is mapped to a unique string with length $w$ and each sample signal is encoded by referring to the string correspondences of each value in the discrete signal and concatenating these strings to construct a long text, \ie corpus.
        
\begin{figure}[t]
        \centering
        \includegraphics[width=0.5\textwidth]{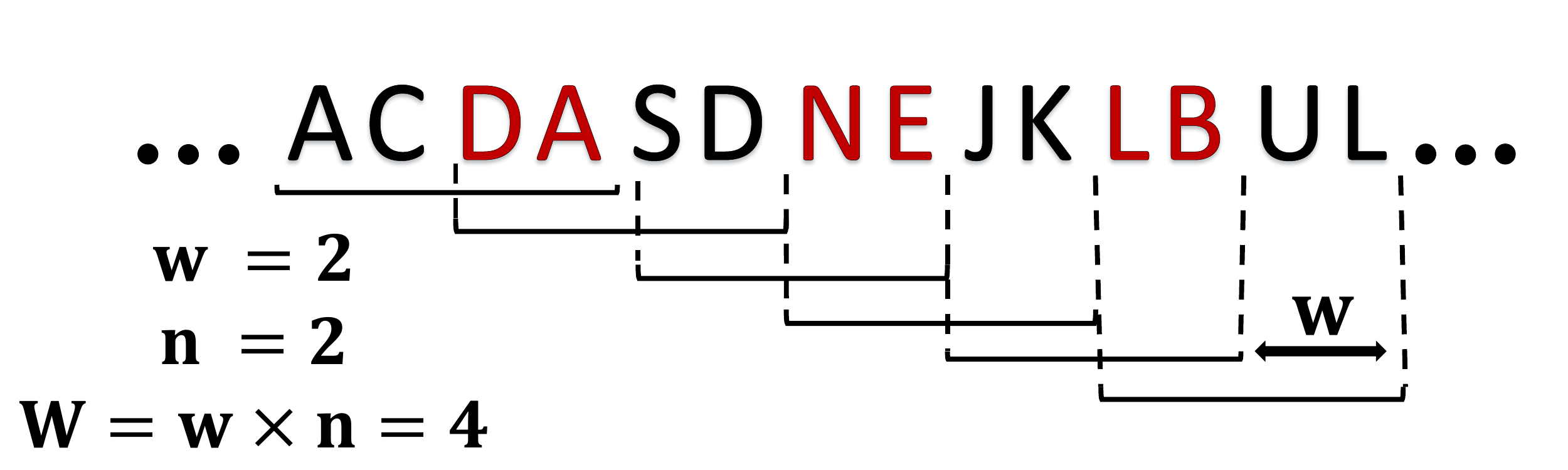}
        \caption{Illustration of bi-gram creation where the word size is $w = 2$ and window size is $W = 4$. }
        \label{figure:n-gram}
    \end{figure}

    \cib{4} To create our \emph{vocabulary}, we consider the corpus created from all encoded signals regardless of labels. Each line in the corpus represents a sample signal, and each word in a line represents the text correspondence of an elevation value in the sample signal. We build a vocabulary from the unique word-based $n$-grams of the document. As illustrated in Figure \ref{figure:n-gram}, a window with size $ W = w \times n $ is slided throughout the corpus and each window content is appended to the vocabulary set. Since the vocabulary set does not contain duplicate entries by definition, we constructed the vocabulary consisting of unique $n$-grams of the given dataset after traversing the corpus by $n$ times with different window sizes. 

    \BfPara{Image-like Representation}
    In image-like transformation, the elevation signals are drawn as line graphs. To draw a line graph, the maximum and minimum values for y-axis are set to be the extremes of each elevation signal, and the lines are colored to encode the value interval in which elevation signal ranges. This method has multiple advantages over other methods---e.g., the alterations of an elevation signal are more visible, and the method results in an efficient utilization of the feature space---which we examined to reach this design choice, but we omit due to lack of space. We use 200 elevation values for each, obtained by dividing the elevation signal into equal-sized parts.
    
    \subsection{Feature Extraction}
    To classify elevation profiles accurately, we extract discriminative features from the elevation profile representations. 
    
    \BfPara{Text-like} In text-like feature extraction, words and non-overlapping occurrences of word sequences are counted, a feature vector for each sample is created with each unique word sequence count being a feature. Finally, the feature vectors are normalized where each feature represents the probability of occurrence of each word in the given sample.
    
    When the dataset is large and diverse, the vocabulary and, consequently, the feature vectors becomes too large and sparse to do computations with. In feature selection phase, in case the length of feature vectors are too long, some rarely occurred features from vocabulary are discarded according to the specified feature frequency threshold. Features are ordered by term frequency across the corpus and the features whose term frequency is under the specified threshold are discarded and a new vocabulary is created. 

    \BfPara{Image-like} \label{Image-like}
    Since we use CNN for images, it is unnecessary to explicitly extract features, since the convolutional layer kernels do that already by learning the filters optimally and efficiently. Therefore, the actual feature extraction mechanism is discussed in the context of classification.
    
    \subsection{Multi-Class Classification}
    For classification, SVM, RFC, MLP, and CNN are used.  
    
    \BfPara{SVM} We use the standard SVM, where the objective is to find the best hyperplane separating classes from one another. We use the linear kernel SVM, which is appropriate for multi-dimensional data~\cite{DBLP:series/faia/2007-160}.  
        
    \BfPara{RFC} We utilized the standard RFC, with 100 trees, and a majority voting is taken over the outcomes of those trees. 
    
    \BfPara{MLP} We use the standard MLP with 100 hidden layers and Adam solver \cite{DBLP:journals/corr/KingmaB14} for weight optimization, since it is claimed to work well for large feature space. MLP is shown to outperform the decision trees~\cite{Lim2000,Eklund02aperformance}.
    
\begin{figure}[t]
        \centering
        \includegraphics[width=0.5\textwidth]{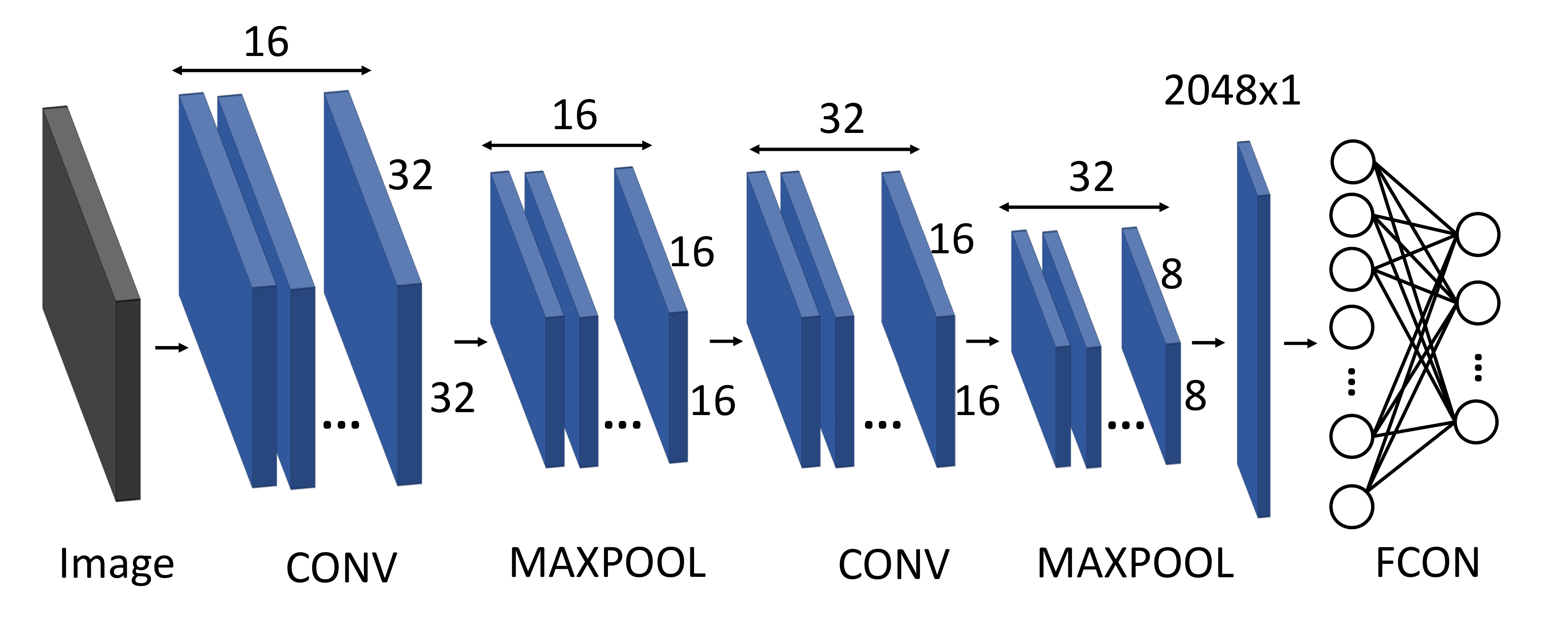}
        \caption{ The CNN architecture used for classification. The input image is passed to a CONV layer. The output is then forwarded to the MAXPOOL layer to fetch the most important feature in a kernel. The data is then passed through following CONV and MAXPOOL layers. The output retrieved from the last MAXPOOL layer is flattened to be passed to a FCON layer whose output is the class probabilities. }
        \label{fig:CNN_architecture}
    \end{figure}

    \BfPara{CNN} Figure \ref{fig:CNN_architecture} illustrates the employed CNN architecture. Two consecutive convolution layers (CONV) are used along with the ReLU activation function and MAX pooling layers (MAXPOOL) before a fully connected layer (FCON). For both of the convolution layers, kernel, stride and padding sizes are decided as 5, 1 and 2 respectively. The distinctive features are selected at the max pooling layers with kernel and stride size of 2, which reduce the dimensions from (32x32) to (8x8) at two passes. The loss is calculated by Cross Entropy Loss function. For optimizing the parameters, the Adam optimizer is used.
            
\section{Evaluation, Results, and Discussion}

We performed evaluations for each dataset and threat models. First, we performed evaluations with text-like representation. For \tma, we performed experiments using 5- and 10-fold cross-validation methods and by fixing the dimension of $n$-grams to 8 on the raw dataset with SVM, MLP and RFC techniques. For \tmb, we tried to find the borough information of a given elevation profile whose city information is known. For this experiment, we use 10-fold cross validation, $n=8$ for the $n$-gram, and SVM, MLP and RFC are employed. For \tmc, we use the same settings as in \tmb. The raw dataset contains overlapped and repetitive samples by nature. In the Simulations subsection, we simulated the same behaviour on the mined datasets and performed the same evaluations for comparison. 

For evaluations using image-like representations, we employed three methods in CNN: unweighted loss function, weighted loss function and fine-tuning. In the unweighted and weighted loss function evaluations, we split the test data from the dataset by considering the sample size of the classes; we assigned probabilities for each class considering the inverse proportion to its size and then randomly select test data with the associated probabilities. In fine-tuning evaluations, we performed a 10-fold validation at the last round where all the classes have the same sample size. 

\subsection{Text-like Data Evaluation}

\BfPara{\cib{1} \tma} We trained and tested models with the raw dataset. As shown in Table \ref{table:raw}, the raw dataset has unbalanced sample size across classes. To mitigate bias, we use the same sample size for each class and change the number of classes at each step. The accuracy results are shown in Table \ref{table:ThreatModel1Results}. Due to the limited number of samples, the accuracy of 4-class classification is lower than 3- and 2-class classification. We observe a higher accuracy with $k$=$10$ than when $k$=$5$ in the $k$-fold cross-validation, perhaps due to the large training data capturing the population's distribution in the first case than in the latter. The results show $ 95.83\% $ accuracy with MLP and 4-class classification. With 3-class classification, we obtained $ 98.51\% $ accuracy with SVM. With 2-class classification, we obtained 98.49\% accuracy with MLP. 

Since the raw dataset is compiled from actual users, exhibiting mobility patterns, about  35\% of the routes are overlapped. In a repetitive and overlapped setting, both training and testing splits may contain similar patterns leading to the high accuracy scores. The results prove that a targeted attack on a person whose activity history is known will be successful with accuracy between 84.80\% and 98.51\%. 

    \begin{table}[t]
        \centering
        \caption{\tma evaluation on raw dataset. Prediction accuracy (\%) with different configurations. 4-class = [{Washington DC, Florida, New York City, California}], 3-class = [{Washington DC, Florida, New York City}], 2-class = [{Washington DC, Florida}]. 5-f: 5-fold cross-validation. 10-f: 10-fold cross-validation. C: the number of classes in the classification. S: sample size of each class. }
        \begin{tabular}{|c|c|c|c|c|c|c|c|}
             \hline
             &  & \multicolumn{2}{c|}{\textbf{SVM}} &          \multicolumn{2}{c|}{\textbf{RFC}} & \multicolumn{2}{c|}{\textbf{MLP}} \\
             C & S & 5-f & 10-f & 5-f & 10-f & 5-f & 10-f \\
             \hline
             2 & 232 & 97.8 & 97.8 &  96.5 & 97.2 & 98.0 & 98.5  \\
             3 & 120 & 98.3 & 98.5 & 96.3 & 97.0 & 97.4 & 97.6  \\
             4 & 18 & 86.8 & 87.5 & 91.0 & 94.4 & 93.0 & 95.8  \\
             \hline
        \end{tabular}
        \label{table:ThreatModel1Results}
    \end{table}

\begin{figure*}[t]
\centering
\begin{subfigure}[Accuracy \label{figure:ThreatModel2ComparisonA}]{\includegraphics[width=0.24\textwidth]{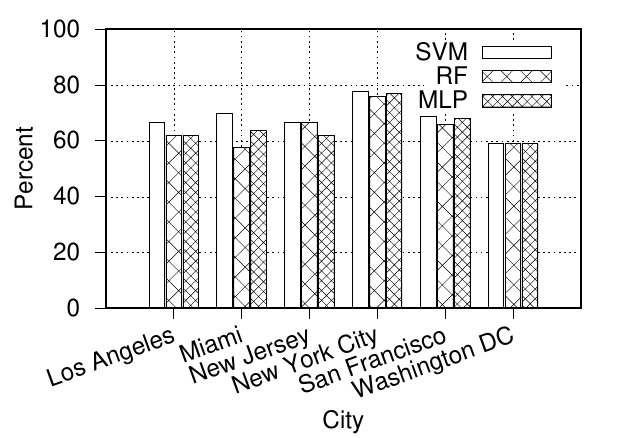}}
\end{subfigure}
\begin{subfigure}[Recall \label{figure:ThreatModel2ComparisonR}]{\includegraphics[width=0.24\textwidth]{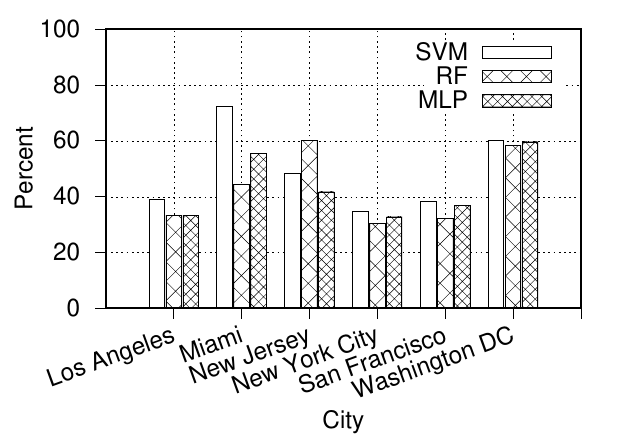}}
\end{subfigure}
\begin{subfigure}[Precision \label{figure:ThreatModel2ComparisonP}]{\includegraphics[width=0.24\textwidth]{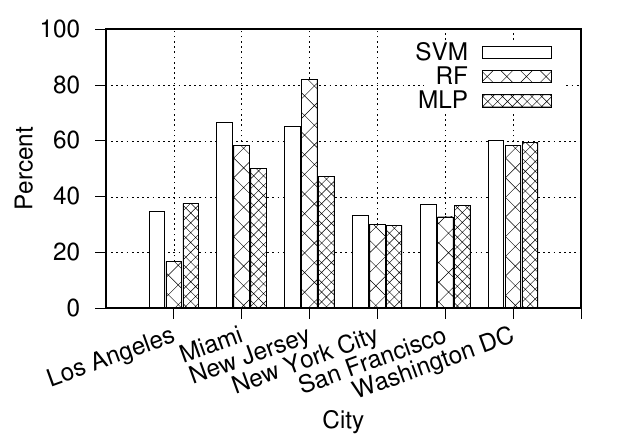}}
\end{subfigure}
\begin{subfigure}[F1 score \label{figure:ThreatModel2ComparisonR}]{\includegraphics[width=0.24\textwidth]{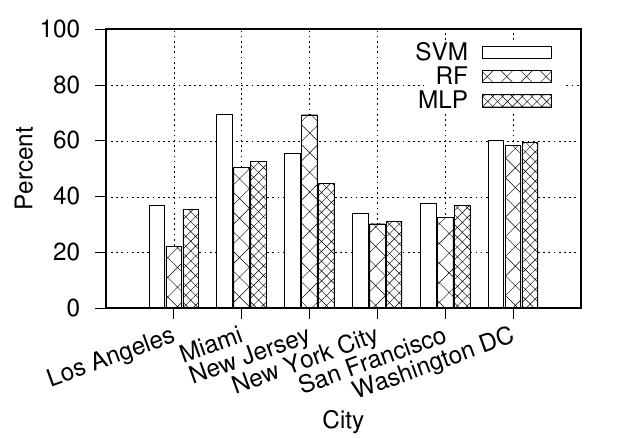}}
\end{subfigure}

\caption{ Accuracy, precision, recall and F1 score of \tmb evaluation with different classification techniques.} 
\label{figure:ThreatModel2Charts}
\end{figure*}

\BfPara{\cib{2} \tmb}
While evaluating \tmb, the borough-level dataset is used. A model is created for each of the cities, Los Angeles, Miami, New Jersey, New York City, San Francisco, and Washington DC, by labeling the data as the name of the corresponding borough and evaluated separately. Figure \ref{figure:ThreatModel2Charts} shows the accuracy, precision, recall and F1 scores of the each model in bar charts. All of the accuracy scores of the models are above 55\% while precision, recall and F1 scores are varying across each model. The main justification for this is that since there is no overlapped or repetitive routes among the mined segments in borough-level dataset, and the elevation differences and elevation sequences are not distinctive enough within a city to decide in which borough is the given test data is. This fact also clarifies the difference between the results of \tma and \tmb. The results of the simulated behaviour will be discussed in the simulations subsection.

\BfPara{\cib{2} \tmc}
In \tmc evaluations, due to sample size differences across the labels in city-level dataset, we followed the same procedure in \tma evaluations. A fixed number of samples was randomly selected from each class for training and testing. Table \ref{table:ThreatModel3Results} shows the results of the evaluation, where we employed 10-fold cross-validation and averaged results of the 10 folds. Per the reported results, we were able to predict the city of an elevation profile among 10 cities with an accuracy of 93.9\%, among 8 cities with an accuracy of 91.93\%, among 7 cities with an accuracy of 90.67\%, among 5 cities with an accuracy of 90.71\%, and among 3 cities with an accuracy of 80.86\%. The success of the city-level estimations when compared to the borough-level estimations is based on the elevation range and sequence differences across cities, which is reasonable, even though the dataset is mined in a similar fashion as in the borough-level dataset. This mining indicates that city-level dataset also does not contain comprehensive, repetitive and overlapped samples. The results of the simulated evaluation will be discussed in the Simulations subsection.

    \begin{table}[t]
        \centering
        \caption{\tmc evaluation on city-level dataset. Prediction accuracy (\textbf{A}), recall (\textbf{R}), F1 score (\textbf{F1}) with different classification techniques and sample size. \textbf{C} column stands for indicating the number of classes in the classification and \textbf{S} column shows sample size of each class.}
        \scalebox{0.75}{
        \begin{tabular}{|c|c|c|c|c|c|c|c|c|c|c|}
             \hline
             &  & \multicolumn{3}{c|}{\textbf{SVM}} & \multicolumn{3}{c|}{\textbf{RFC}} & \multicolumn{3}{c|}{\textbf{MLP}} \\
             \textbf{C} & \textbf{S} & \textbf{A} & \textbf{R} & \textbf{F1} & \textbf{A} & \textbf{R} & \textbf{F1} & \textbf{A} & \textbf{R} & \textbf{F1} \\
             \hline
             3 & 743 & 80.0 & 69.8 & 70.2 & 
                79.1 & 68.4 & 68.4 & 
                80.9 & 71.2 & 71.6 \\
             5 & 362 & 90.7 & 77.7 & 78.4 & 
                89.4 & 74.8 & 76.0 & 
                90.5 & 77.4 & 78.4 \\
             7 & 266 & 90.7 & 66.7 & 66.5 & 
                89.0 & 61.1 & 61.0 & 
                90.0 & 64.3 & 64.4 \\
             8 & 155 & 91.9 & 68.6 & 68.5 & 
                88.9 & 57.0 & 60.3 & 
                90.9 & 65.1 & 64.5 \\
             10 & 82 & 93.9 & 70.2 & 70.4 & 
                92.4 & 58.1 & 57.5 & 
                92.9 & 63.7 & 63.3 \\
             \hline
        \end{tabular}
        }
        \label{table:ThreatModel3Results}
    \end{table}

\subsubsection{Simulations}
The mined datasets do not contain overlapped or duplicate samples as in the raw dataset. In this evaluation, we simulated overlapped mined datasets and performed evaluations under the same threat models. 

\begin{figure*}[t]
\centering
\begin{subfigure}[Accuracy \label{figure:ThreatModel2ComparisonA}]{\includegraphics[width=0.24\textwidth]{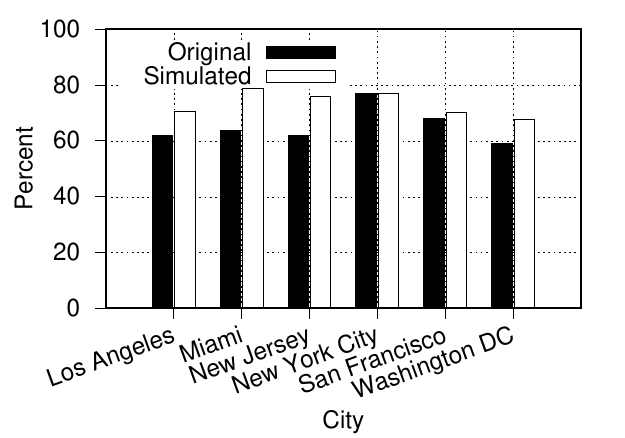}}
\end{subfigure}
\begin{subfigure}[Recall \label{figure:ThreatModel2ComparisonR}]{\includegraphics[width=0.24\textwidth]{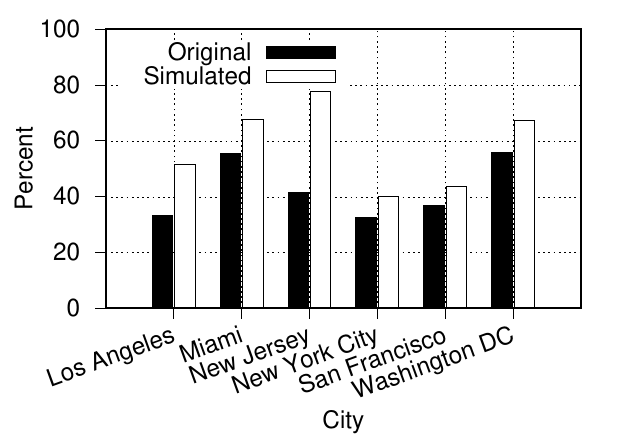}}
\end{subfigure}
\begin{subfigure}[Precision \label{figure:ThreatModel2ComparisonP}]{\includegraphics[width=0.24\textwidth]{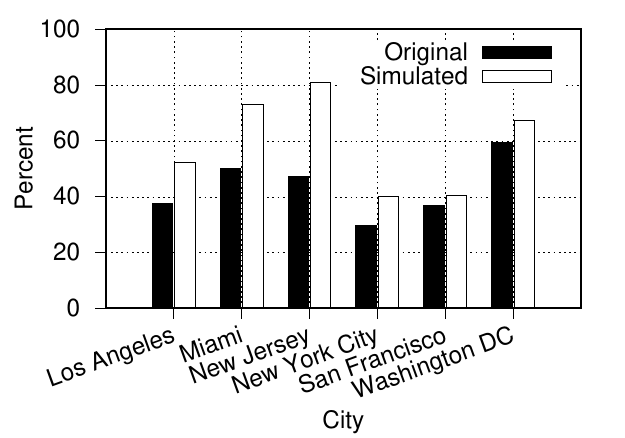}}
\end{subfigure}
\begin{subfigure}[F1 score \label{figure:ThreatModel2ComparisonR}]{\includegraphics[width=0.24\textwidth]{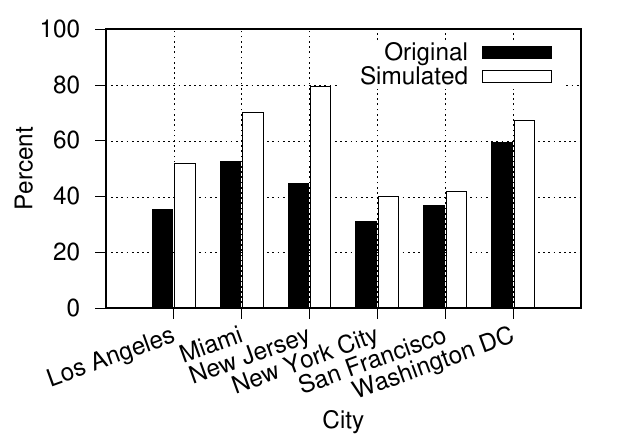}}
\end{subfigure}
\vspace{-4mm}
\caption{ Comparison of the \tmb simulated dataset evaluation accuracy scores by using Multi-Layer Perceptron classification and the original dataset evaluation accuracy scores by using Multi-Layer Perceptron classification.} 
\label{figure:ThreatModel2Comparison}
\end{figure*}

\BfPara{Simulation of \tmb} For the city-level estimation evaluations, we rebuilt a simulation dataset with 30 - 34\% overlap ratio for each region within the cities. The same evaluation procedures are then followed as the original mined dataset, which is 10-fold cross validation with fixed $n$-grams size of 8. Figure \ref{figure:ThreatModel2Comparison} shows comparison between the results of MLP classification, confirming our previous hypothesis that having overlapped route samples would increase the accuracy. Since the mined dataset is not specific to any target user's behaviour, it is anticipated to result in less accuracy than the \tma evaluation accuracy scores.

\BfPara{Simulation of \tmc} For \tmc's simulated evaluations, we rebuilt a simulation dataset with 35\% overlap ratio for each city and performed the same evaluation with 10-fold cross validation and 8-grams. Table \ref{table:ThreatModel3SimulationResults} shows the results. By comparing Table \ref{table:ThreatModel3Results} to Table \ref{table:ThreatModel3SimulationResults}, we notice that the accuracy, recall, precision, and F1 scores have increased for all classification techniques as expected. The improvements prove our previous hypothesis that having similar patterns in a dataset affects the success of the attack.
 
    \begin{table}[t]
        \centering
        \caption{ \tmc evaluation on city-level dataset  when 35\% overlap is introduced. Prediction accuracy (\textbf{A}), recall (\textbf{R}), F1 score (\textbf{F1}) with different configurations. \textbf{C} column stands for indicating the number of classes in the classification and \textbf{S} column shows sample size of each class.}
        \scalebox{0.75}{
        \begin{tabular}{|c|c|c|c|c|c|c|c|c|c|c|}
             \hline
             \multicolumn{2}{|c|}{} & \multicolumn{3}{c|}{\textbf{SVM}} & \multicolumn{3}{c|}{\textbf{RFC}} & \multicolumn{3}{c|}{\textbf{MLP}} \\
             \textbf{C} & \textbf{S} & \textbf{A} & \textbf{R} & \textbf{F1} & \textbf{A} & \textbf{R} & \textbf{F1} & \textbf{A} & \textbf{R} & \textbf{F1} \\
             \hline
             3 & 966 & 91.7 & 82.7 & 82.8 & 
                89.0 & 77.8 & 79.1 & 
                92.4 & 84.0 & 84.1 \\
             5 & 470 & 94.6 & 81.6 & 81.2 & 
                93.7 & 78.7 & 78.4 & 
                95.6 & 85.0 & 84.7 \\
             7 & 338 & 93.6 & 72.1 & 72.5 & 
                92.4 & 68.4  & 68.8 & 
                93.9 & 73.4  & 73.4 \\
             8 & 202 & 94.7 & 75.4 & 74.9 & 
                93.2 & 67.8  & 66.9 & 
                94.6 & 74.9  & 74.2 \\
             10 & 107 & 94.4 & 71.4 & 72.5 & 
                93.6 & 67.7  & 66.9 & 
                93.6 & 68.9  & 69.8 \\
             \hline
            \end{tabular}
            }
        \label{table:ThreatModel3SimulationResults}
    \end{table}
    
\subsection{Image-like Data Evaluations}

    \BfPara{Dealing with Unbalanced Dataset}
    The original datasets are unbalanced and there are various methods to deal with unbalanced datasets, including downsampling, oversampling, and creating synthetic samples from existing ones. Among these methods, downsampling and oversampling are the easiest ones to explore, although downsampling leads to losing great amount of data and oversampling rises the chances of getting lower accuracy as the misclassified duplicated samples increases the false ratio.
    
    \BfPara{Weighted Loss Function} For the unbalanced dataset, we utilized a weighted loss function while training the CNN and used all the data in the dataset. By assigning a class weight that is inversely proportional to the sample size of the class, we signify samples of small classes while calculating the loss, thus their effect does not easily wear off. 
    
\begin{figure}[t]
        \centering
        \includegraphics[width=0.5\textwidth]{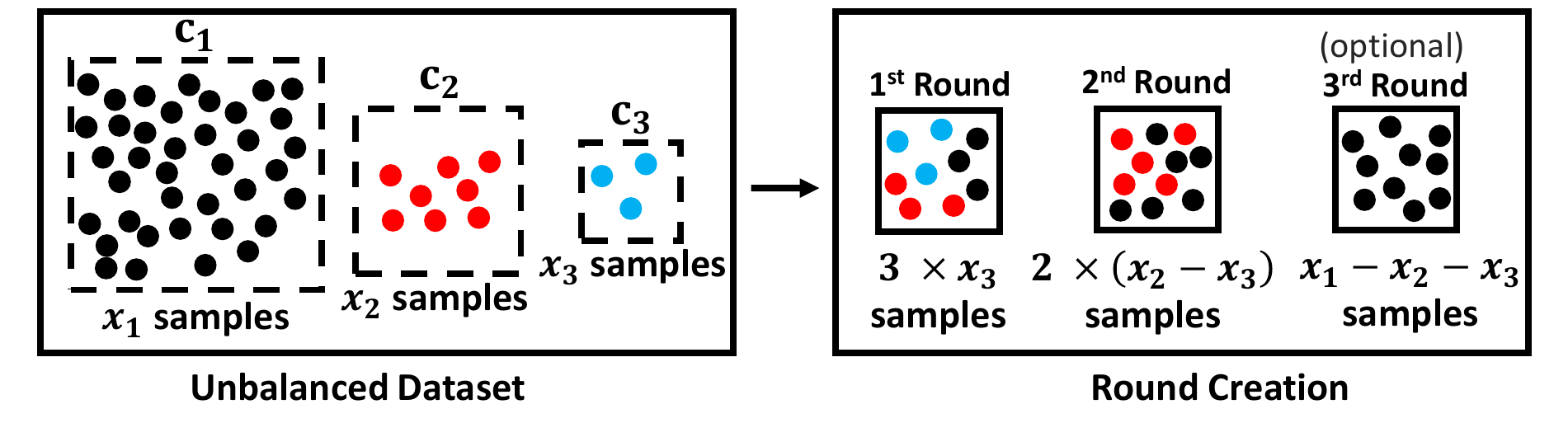}
        \caption{An illustration of round creation from an unbalanced dataset of 3 classes. }
        \label{figure:round_creation}
    \end{figure}

\begin{figure}[t]
        \centering
        \includegraphics[width=0.5\textwidth]{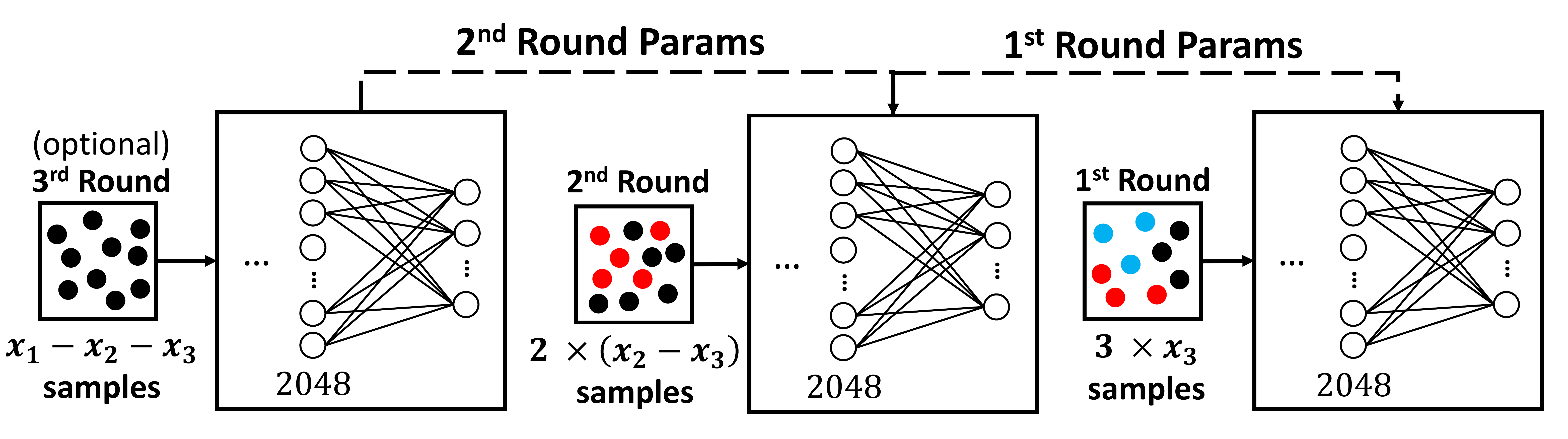}
        \caption{An illustration of the fine-tuning pipeline for an unbalanced dataset of 3 classes.}
        \label{figure:fine-tuning}
    \end{figure}

    \BfPara{Fine-Tuning with Different Samples} Fine-tuning is a common technique in deep learning, and is used for re-training a complex pretrained model with another dataset. To address the unbalanced dataset, we take advantage of fine-tuning in a different manner. Namely, we introduced rounds and created a set of small datasets from the unbalanced datasets for each round. As illustrated in Figure \ref{figure:round_creation}, several small and balanced datasets are created by randomly selecting samples. For each consecutive rounds, samples of one or more classes are discarded, and the round dataset is  created from the remaining classes. After round dataset creation, the model is trained with the round dataset that contains \emph{the least number of classes}, \ie the lattermost created round dataset. At each step, the model is  re-trained using the same or different hyperparameters until all the rounds expire. The dataset ordering of the rounds are reversed, since the impact of the smallest dataset would wear off if the model is trained with the same order of round dataset creation, which conflicts with the whole idea. As illustrated in Figure \ref{figure:fine-tuning}, while re-training, the parameters of the previous model are passed to the model of the next round. The hyperparameters of each round can be tuned accordingly. For instance, for the last round, where we include all of the classes, the learning rate can be reduced in order to find the loss minima.
            
    To evaluate our attacks on the image-like data, the elevation profiles are converted into a dataset of images and rounds using the configurations and steps discussed above. Table \ref{table:Comparison} highlights the maximum achieved prediction accuracy along with comparisons among methods. 
 
    \begin{table}[t]
        \centering
        \caption{Comparison of maximum achieved accuracy across different methods. The Unweighted Loss(UW) column is not considered while deciding the maximum accuracy, as the results are biased. The maximum accuracy of each evaluation is written \textbf{bold}, the results that are not considered are written \textit{italic}. Location abbreviations are in Table~\ref{table:borough-level}.}
        \begin{tabular}{|l|c|c|c|c|}
             \hline
             & \multicolumn{1}{c|}{\textbf{Text-like}} & \multicolumn{3}{c|}{\textbf{Image-like}} \\
             
             Methods & DS & \makecell{UWL \\ \textit{(biased)}} & WL & FT \\
             \hline
             \tma & \textbf{95.83} & \textit{96.98} & 95.23 & 87.93 \\
             \hline
             \tmb: LA   & 65.13 & \textit{68.85}     & \textbf{68.39}   & 63.63 \\
             \tmb: MIA  & 68.65 & \textit{88.96}     & \textbf{86.80}   & 62.50 \\
             \tmb: NJ   & 63.52 & \textit{93.45}     & \textbf{79.42}   & 57.14 \\
             \tmb: NYC  & 78.85 & \textit{74.20}     & \textbf{79.37}   & 72.79 \\
             \tmb: SF   & 64.52 & \textit{67.20}     & \textbf{78.70}   & 65.38 \\
             \tmb: WDC  & 60.79 & \textit{62.79}     & 70.28            & \textbf{71.50} \\
             \hline
             \tmc & \textbf{93.90} & \textit{92.51} & 92.82 & 89.00 \\
             \hline
        \end{tabular}
        \label{table:Comparison}
    \end{table}

\BfPara{Weighted vs. Unweighted Loss Function} 
To observe the impact of the weighted loss function, we conducted evaluations without giving any weight to the classes in the loss function while using an unbalanced dataset. We note that the unweighted loss function evaluation results are biased due to the unbalanced dataset. Table \ref{table:Comparison} shows the maximum achieved accuracy for each dataset and method. Even though the weighted loss function evaluation results are biased, which \emph{seems} successful in outputting the largest class used during training and testing, the biased results remain behind 4 evaluations out of 8. For the remaining evaluations, and \textit{except \tmb: NJ}, the difference between the results is insignificant. When the biased results are excluded for \tmb evaluations, the weighted loss function performed better than text-like and fine-tuning methods. However, in \tma and \tmc, the accuracy scores between methods are considerably close. Thus, we conclude the weighted loss function improved the prediction performance primarily for \tmb.

\BfPara{Fine-tuning} Round datasets are created from the original data. For \tma, with 4 classes, 3 rounds are created. For \tmc, with 10 classes, 5 rounds were created by eliminating 1, 2, 1 and 2 classes at each round, respectively. The dataset of \tmb can be considered as a compilation of the dataset of 6 cities: Los Angeles (3 rounds), Miami (3 rounds), New Jersey (2 rounds), New York City (4 rounds), San Francisco (2 rounds), and Washington DC (1 round). Even though the main idea is using all the data we have, we decided to downsample the classes with large sample size. For instance, in the evaluation of \tmb: New York City, the biggest class has 5,455 samples where the second biggest class has 960 samples. In such cases, we did not create additional round for only one class as this round would have strong influence over the predictions, \ie may cause overfitting.

Table \ref{table:Comparison} shows the fine-tuning method outperformed other methods only for \tmb: Washington DC. The difference between the fine-tuning evaluation of Washington DC and others is that we were able to create only one round from the data in the former. Overall, the fine-tuning evaluation is not as successful as the weighted loss function evaluation, since we still lost some data while creating rounds.

    \begin{table}[t]
        \centering
        \caption{ The fine-tuning results for \tma and \tmc as the epoch size changes. }
        \begin{tabular}{|c|c|c|c|c|c|c|}
             \hline
             & \multicolumn{3}{c|}{\tma} & \multicolumn{3}{c|}{\tmc} \\
             \textbf{Epoch Size} & \textbf{500} & \textbf{1000} & \textbf{2000} & \textbf{500} & \textbf{1000} & \textbf{2000} \\
             \hline
             Accuracy & 79.3 & 87.9 & 82.7 & 86.0 & 89.0 & 87.8 \\
             Recall & 55.8 & 67.5 & 63.1 & 29.7 & 45.3 & 38.9  \\
             Specificity & 86.3 & 92.6 & 88.4 & 92.2 & 93.9 & 93.2 \\
             F1 Score & 58.6 & 68.2 & 63.3 & 36.2 & 45.4 & 41.1 \\
             \hline
        \end{tabular}
        \label{table:Fine-tune_results_1_3}
    \end{table}
    
    \begin{table}[t]
        \centering
        \caption{ The fine-tuning results for \tmb as the epoch size is 1000 and learning rate is 0.001 for all rounds. }
        \begin{tabular}{|c|c|c|c|c|c|c|}
             \hline
             \textbf{} & \textbf{LA} & \textbf{MIA} & \textbf{NJ} & \textbf{NYC} & \textbf{SF} & \textbf{WDC} \\
             \hline
             Accuracy & 63.6 & 62.5 & 57.1 & 72.8 & 65.4 & 71.5 \\
             Recall & 28.0 & 25.6 & 40.0 & 18.1 & 30.7 & 73.2  \\
             Specificity & 75.8 & 75.9 & 66.7 & 83.4 & 76.3 & 73.2 \\
             F1 Score & 28.8 & 28.6 & 37.5 & 18.4 & 31.4 & 73.4 \\
             \hline
        \end{tabular}
        \label{table:Fine-tune_results_2}
    \end{table}

\section{Related Work}
In this work, we addressed \emph{location privacy} in activity trackers  using side channel information from publicly shared elevation profile, a topic that is related to various pieces of the literature. In the following, we review some of those studies.

Most location privacy breaches are caused since users do not know why or how to preserve location privacy. \citet{aktypi2017a} developed a tool to examine possible privacy exposures of users in their social networks where the data is mostly collected from wearable devices. Using this tool, the authors aimed to enhance the awareness of information leakage in social networks, particularly fitness apps in which the data retrieved from wearable devices is shared on social networks. \citet{Abdelmoty2017} aimed to increase awareness of location privacy on geo-social networks by surveying 186 users, where 77\% of them indicated they use location-based services often, several times a day, and 47\% of them reported that they were not aware that the location-based apps collect and store location information when users select the private location option. Moreover, 43\% of respondents were not aware that application may share the location information with third parties.

Despite the methods employed to preserve location privacy, several attacks are devised to uncover supposedly protected locations. Experiments for revealing exact locations from trajectories with private zones are conducted on Strava \cite{217618}. Researchers found the exact end points associated with users, even when such users selected the private zone option when sharing the training route. In another study, location trajectories of users are recovered from publicly available aggregated mobility data obtained from GSM operators \cite{8356232}. The attack relies on tracking the regularity---i.e., coming across the same location trace in the aggregated data regularly---and uniqueness---i.e., the location trace belongs to a unique user---of the user mobility traces to recover trajectories.

As our study exemplifies, online social networks lays under the scope of privacy breach risks for users. \citet{DBLP:conf/icwsm/ZhengHYKL15} shows that sharing data which reveals spatiotemporal features of users' mobility patterns on online social networks reveal sensitive information such as home location, using a different form of data, \ie multimedia. \citet{DBLP:conf/icwsm/0004WSM15} shows that location-based social networks are vulnerable also to identity privacy breaches by revealing the identity of users by observing their mobility patterns.

Several attacks against general location privacy methods are proposed~\cite{Wernke2014}. The homogeneity attack ~\cite{1617392} is an attack on k-anonymity to infer data of interest from other shared data. \citet{1617392}  illustrated a scenario where an adversary infers the illness of a target person from available information, the zip code, age, etc. The same method can be applied to infer location data. In location distribution attacks \cite{4417152}, the adversary exploits the fact that users are mostly not uniformly distributed in the location space. Another attack \cite{5958033} utilized the aggregated traffic statistics and environmental context information. The attack scenario includes an adversary who tries to reveal the possible location of the target by making use of the fact that the probability of target's whereabouts is not uniformly distributed. Map matching methods \cite{10.1007/978-3-540-72037-9_8} aim to restrict the obfuscated area to a smaller but plausible area by removing irrelevant areas. Movement boundary attacks were explored~\cite{Ghinita:2009:PVL:1653771.1653807}, where the adversary aims to calculate the movement boundary of a target by chasing the position queries and updates of the target. After calculating the boundary, the location of interest, such as home or work place, is inferred and the irrelevant locations are discarded.

Although we did not directly touch upon preserving the location privacy in our study, there has been a few related studies in this space. The fast-growing need of preserving location privacy over the aforementioned attacks excited researchers' attention. Researchers introduce obfuscation methods such as decreasing the quality of the location by introducing inaccuracy and imprecision \cite{Duckham:2005:FMO:2154273.2154286}. Additionally, the term k-anonymity is defined as obscuring the location information of individuals with \emph{k} number of other individuals within the region \cite{Sweeney:2002:AKA:774544.774553,Gkoulalas-Divanis:2010:PKL:1882471.1882473}.

\section{Conclusion}
In this paper, we presented new attacks on location privacy using only elevation profiles publicly available from fitness trackers. The attacks are categorized into three types: predicting location by knowing the activity history of the target, predicting the borough by knowing the city of the target, and predicting the city of the target without any prior knowledge. The key contributions of our work are proving the concept that hiding the route of a workout and sharing only the elevation profile is not sufficient to preserve location privacy, defining a new attack surface by creating scenarios for possible threat models, and providing a machine-learning approach to realize such threat into attacks.

To validate our attacks we created three datasets by collecting data from Strava users, and mining data using Strava and Google Elevation API. We preprocessed the datasets by employing Natural Language Processing and Computer Vision approaches, and then employed classification techniques to predict the location from elevation profiles. En route, we defined three threat models and evaluated each of them individually on the different datasets. As a result of the evaluations, we were able to identify the corresponding location of an elevation profile with accuracy between 59.59\% and 95.83\%. While a simple defense, such hiding elevation would work, it might reduce the usability of the fitness applications. In the future, we will explore compatible defenses.

\bibliographystyle{aaai}
\bibliography{ref}

\begin{thebibliography}{}

\bibitem[\protect\citeauthoryear{Abdelmoty and Alrayes}{2017}]{Abdelmoty2017}
Abdelmoty, A.~I., and Alrayes, F.
\newblock 2017.
\newblock Towards understanding location privacy awareness on geo-social
  networks.
\newblock {\em ISPRS International Journal of Geo-Information} 6(4).

\bibitem[\protect\citeauthoryear{Aktypi, Nurse, and
  Goldsmith}{2017}]{aktypi2017a}
Aktypi, A.; Nurse, J.; and Goldsmith, M.
\newblock 2017.
\newblock Unwinding ariadne's identity thread: Privacy risks with fitness
  trackers and online social networks.
\newblock  1--11.
\newblock Association for Computing Machinery.

\bibitem[\protect\citeauthoryear{Duckham and
  Kulik}{2005}]{Duckham:2005:FMO:2154273.2154286}
Duckham, M., and Kulik, L.
\newblock 2005.
\newblock A formal model of obfuscation and negotiation for location privacy.
\newblock In {\em Proceedings of the Third International Conference on
  Pervasive Computing}, PERVASIVE'05,  152--170.
\newblock Berlin, Heidelberg: Springer-Verlag.

\bibitem[\protect\citeauthoryear{Eklund}{2002}]{Eklund02aperformance}
Eklund, P.~W.
\newblock 2002.
\newblock A performance survey of public domain supervised machine learning
  algorithms.
\newblock Technical report.

\bibitem[\protect\citeauthoryear{Friedland and Sommer}{2010}]{Cybercasing}
Friedland, G., and Sommer, R.
\newblock 2010.
\newblock Cybercasing the joint: On the privacy implications of geo-tagging.
\newblock  1--8.

\bibitem[\protect\citeauthoryear{Ghinita \bgroup et al\mbox.\egroup
  }{2009}]{Ghinita:2009:PVL:1653771.1653807}
Ghinita, G.; Damiani, M.~L.; Silvestri, C.; and Bertino, E.
\newblock 2009.
\newblock Preventing velocity-based linkage attacks in location-aware
  applications.
\newblock In {\em Proceedings of the 17th ACM SIGSPATIAL International
  Conference on Advances in Geographic Information Systems}, GIS '09,
  246--255.
\newblock New York, NY, USA: ACM.

\bibitem[\protect\citeauthoryear{Gkoulalas-Divanis, Kalnis, and
  Verykios}{2010}]{Gkoulalas-Divanis:2010:PKL:1882471.1882473}
Gkoulalas-Divanis, A.; Kalnis, P.; and Verykios, V.~S.
\newblock 2010.
\newblock Providing k-anonymity in location based services.
\newblock {\em SIGKDD Explor. Newsl.} 12(1):3--10.

\bibitem[\protect\citeauthoryear{Hassan, Hussain, and Bates}{2018}]{217618}
Hassan, W.~U.; Hussain, S.; and Bates, A.
\newblock 2018.
\newblock Analysis of privacy protections in fitness tracking social networks
  -or- you can run, but can you hide?
\newblock In {\em 27th {USENIX} Security Symposium ({USENIX} Security 18)},
  497--512.
\newblock Baltimore, MD: {USENIX} Association.

\bibitem[\protect\citeauthoryear{Kingma and
  Ba}{2014}]{DBLP:journals/corr/KingmaB14}
Kingma, D.~P., and Ba, J.
\newblock 2014.
\newblock Adam: {A} method for stochastic optimization.
\newblock {\em CoRR} abs/1412.6980.

\bibitem[\protect\citeauthoryear{Krumm}{2007}]{10.1007/978-3-540-72037-9_8}
Krumm, J.
\newblock 2007.
\newblock Inference attacks on location tracks.
\newblock In LaMarca, A.; Langheinrich, M.; and Truong, K.~N., eds., {\em
  Pervasive Computing},  127--143.
\newblock Berlin, Heidelberg: Springer Berlin Heidelberg.

\bibitem[\protect\citeauthoryear{Lim, Loh, and Shih}{2000}]{Lim2000}
Lim, T.-S.; Loh, W.-Y.; and Shih, Y.-S.
\newblock 2000.
\newblock A comparison of prediction accuracy, complexity, and training time of
  thirty-three old and new classification algorithms.
\newblock {\em Machine Learning} 40(3):203--228.

\bibitem[\protect\citeauthoryear{Loughran}{2019}]{AdvancedDeanonymizationThroughStrava}
Loughran, S.
\newblock 2019.
\newblock Advanced deanonymization through strava.
\newblock
  \url{http://steveloughran.blogspot.com/2018/01/advanced-denanonymization-through-strava.html}.
\newblock Accessed: 2019-03-20.

\bibitem[\protect\citeauthoryear{{Machanavajjhala} \bgroup et al\mbox.\egroup
  }{2006}]{1617392}
{Machanavajjhala}, A.; {Gehrke}, J.; {Kifer}, D.; and {Venkitasubramaniam}, M.
\newblock 2006.
\newblock L-diversity: privacy beyond k-anonymity.
\newblock In {\em 22nd International Conference on Data Engineering (ICDE'06)},
   24--24.

\bibitem[\protect\citeauthoryear{Maglogiannis \bgroup et al\mbox.\egroup
  }{2007}]{DBLP:series/faia/2007-160}
Maglogiannis, I.; Karpouzis, K.; Wallace, M.; and Soldatos, J., eds.
\newblock 2007.
\newblock {\em Emerging Artificial Intelligence Applications in Computer
  Engineering - Real Word {AI} Systems with Applications in eHealth, HCI,
  Information Retrieval and Pervasive Technologies}, volume 160 of {\em
  Frontiers in Artificial Intelligence and Applications}. {IOS} Press.

\bibitem[\protect\citeauthoryear{{Mokbel}}{2007}]{4417152}
{Mokbel}, M.~F.
\newblock 2007.
\newblock Privacy in location-based services: State-of-the-art and research
  directions.
\newblock In {\em 2007 International Conference on Mobile Data Management},
  228--228.

\bibitem[\protect\citeauthoryear{P~Higgins}{2015}]{higgins}
P~Higgins, J.
\newblock 2015.
\newblock Smartphone applications for patients’ health \& fitness.
\newblock {\em The American journal of medicine} 129.

\bibitem[\protect\citeauthoryear{Polakis \bgroup et al\mbox.\egroup
  }{2015}]{Polakis:2015:WWP:2810103.2813605}
Polakis, I.; Argyros, G.; Petsios, T.; Sivakorn, S.; and Keromytis, A.~D.
\newblock 2015.
\newblock Where's wally?: Precise user discovery attacks in location proximity
  services.
\newblock In {\em Proceedings of the 22Nd ACM SIGSAC Conference on Computer and
  Communications Security}, CCS '15,  817--828.
\newblock New York, NY, USA: ACM.

\bibitem[\protect\citeauthoryear{Rossi \bgroup et al\mbox.\egroup
  }{2015}]{DBLP:conf/icwsm/0004WSM15}
Rossi, L.; Williams, M.~J.; Stich, C.; and Musolesi, M.
\newblock 2015.
\newblock Privacy and the city: User identification and location semantics in
  location-based social networks.
\newblock In {\em Proceedings of the Ninth International Conference on Web and
  Social Media, {ICWSM} 2015, University of Oxford, Oxford, UK, May 26-29,
  2015},  387--396.

\bibitem[\protect\citeauthoryear{{Shokri} \bgroup et al\mbox.\egroup
  }{2011}]{5958033}
{Shokri}, R.; {Theodorakopoulos}, G.; {Le Boudec}, J.; and {Hubaux}, J.
\newblock 2011.
\newblock Quantifying location privacy.
\newblock In {\em 2011 IEEE Symposium on Security and Privacy},  247--262.

\bibitem[\protect\citeauthoryear{Sweeney}{2002}]{Sweeney:2002:AKA:774544.774553}
Sweeney, L.
\newblock 2002.
\newblock Achieving k-anonymity privacy protection using generalization and
  suppression.
\newblock {\em Int. J. Uncertain. Fuzziness Knowl.-Based Syst.} 10(5):571--588.

\bibitem[\protect\citeauthoryear{{Tu} \bgroup et al\mbox.\egroup
  }{2018}]{8356232}
{Tu}, Z.; {Xu}, F.; {Li}, Y.; {Zhang}, P.; and {Jin}, D.
\newblock 2018.
\newblock A new privacy breach: User trajectory recovery from aggregated
  mobility data.
\newblock {\em IEEE/ACM Transactions on Networking} 26(3):1446--1459.

\bibitem[\protect\citeauthoryear{Wernke \bgroup et al\mbox.\egroup
  }{2014}]{Wernke2014}
Wernke, M.; Skvortsov, P.; D{\"u}rr, F.; and Rothermel, K.
\newblock 2014.
\newblock A classification of location privacy attacks and approaches.
\newblock {\em Personal and Ubiquitous Computing} 18(1):163--175.

\bibitem[\protect\citeauthoryear{Zheng \bgroup et al\mbox.\egroup
  }{2015}]{DBLP:conf/icwsm/ZhengHYKL15}
Zheng, D.; Hu, T.; You, Q.; Kautz, H.~A.; and Luo, J.
\newblock 2015.
\newblock Towards lifestyle understanding: Predicting home and vacation
  locations from user's online photo collections.
\newblock In {\em Proceedings of the Ninth International Conference on Web and
  Social Media, {ICWSM} 2015, University of Oxford, Oxford, UK, May 26-29,
  2015},  553--561.

\end{thebibliography}

\end{document}